# Enhancing Multi-Agent Based Simulation with Human-Agents Interactive Spatial Behaviour

Yee Ming Chen, Bo-Yuan Wang and Hung-Ming Shiu

**Abstract**—We are exploring the enhancement of models of agent behaviour with more "human-like" decision making strategies than are presently available. Our motivation is to developed with a view to as the decision analysis and support for electric taxi company under the mission of energy saving and reduction of CO2, in particular car-pool and car-sharing management policies. In order to achieve the object of decision analysis for user, we provide a human-agents interactive spatial behaviour to support user making decision real time. We adopt passenger average waiting time and electric taxi average idle time as the performance measures and decision support fro electric taxi company. Finally, according to the analysis result, we demonstrate that our multi-agent simulation and GUI can help users or companies quickly make a quality and accurate decision to reduce the decision-making cost and time.

**Index Terms**—Human spatial behaviors , Multi-agent , Simulation.

## 1 INTRODUCTION

Simulations help to view the same phenomena at different levels of abstraction and hence aid in easy understanding for various users who have different knowledge of the stratagies under consideration. They are also useful where the effects of implementation of a transporation system are difficult to predict and the actual planning is quite costly. Such simulations can be used to evaluate modifications not only under nominal conditions, but also under hypothetical scenarios that would be difficult to observe in the real world.

We all know that mass transportation can significantly reduce the emission of $CO_2$ and save a lot of energy consuming. But, in order to reduce the operation cost, mass transportation abandons flexibility and efficiency. Taxicabs play an important role in bridging the gap between private transportation, buses, and rail systems in urban areas[1,2]. However, taxi still cause the generation of $CO_2$ and lot of energy consuming. In order to improve the condition plus the evolution of power, electric taxi comes with the tide of fashion. Generally speaking, electric taxi has to go around a city for searching passengers. This service mode causes lot of electricity wasting during the idle time (for searching passengers). Dial-a-ride (DAR) is another service mode for electric taxi fleet[3]. Electric taxi drivers don't need to go around a city to search passengers. They just wait for an order delivered from control center and then go to the designated location perform the service. So electric taxi company applies dial-a-ride service mode also can economize the energy consuming and reduce the passenger waiting time. And passenger will spend a lot of time to wait for an electric taxi. Meanwhile, electric taxi still consumes electricity during operation period. When an electric taxi lacks of electricity, it can't serve any passenger. And electricity consuming during the idle time will become energy wasting. Therefore, electric stations establishment and reduction of electric taxi idle time are very important issues. In order to deal with these issues, we have to know how to reduce electric idle time and estabilsh electric stations. For reduction of electric idle time, the solutions is establishing a small electric taxi fleet. Nonetheless, insufficient electric taxi causes higher passenger waiting time. Contrary, an excess of electric taxi causes higher electric taxi idle time. Hence, how to balance the passenger waiting time and electric taxi idle time is a very important issue. For our research, in order to understand the development of electric taxi operation system, we apply multi-agent simulation in interactive spatial behaviour analysis and decision-making processes in the field of transportation planning. Therefore, the purposes of this research are listed as follows:

1. To manage and observe the behaviors of electric taxi DAR operation system.
2. To analyze and discuss the simulation result of electric taxi DAR operation system with car-pool and car-sharing.
3. To construct graphic user interface (GUI) to connect with agents and users to validate and the interactive spatial behaviour analysis on electric taxi DAR operation system.

The remainder of our paper is structured as follows: In section 2, we proposed the framework of multi-agent and agents' interactive spatial behaviour. Section 3 describes electric-taxi Dial-a-ride (DAR) operation system and evaluate electric taxi fleet scale and its management policies (paths planning, Car-Pool and Car-Sharing) to observe the result of simulation. Finally, section 5 draws conclu-

- *Yee Ming Chen is with the Department of Industrial Engineering and Management, Yuan Ze University, Taoyuan, Taiwan.*
- *Bo-Yuan Wang, Hung-Ming Shiu are with the Department of Industrial Engineering and Management, Yuan Ze University, Taoyuan, Taiwan.*

sions, and points out future research.

## 2 THE FRAMEWORK OF MULTI-AGENT

The framework of multi-agent is diagramed as in figure 1 . Container A, B and C represent three different types of agents. Each type of container includes many agents. Every agent has his own goal and can interact with each other. On the other hand, user can interact with agents via graphic user interface (GUI).

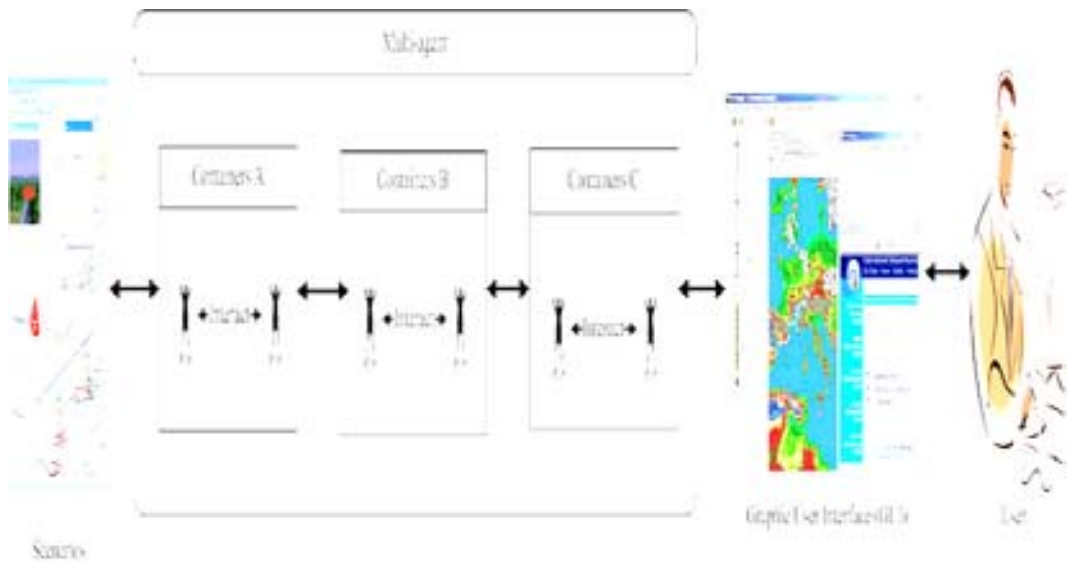

Fig. 1. The framework of Multi-agent

Three types of container have six kinds of agents: electric-taxis, control center, passengers, electric stations, road and stops, which are located in these containers as the and figure 2.

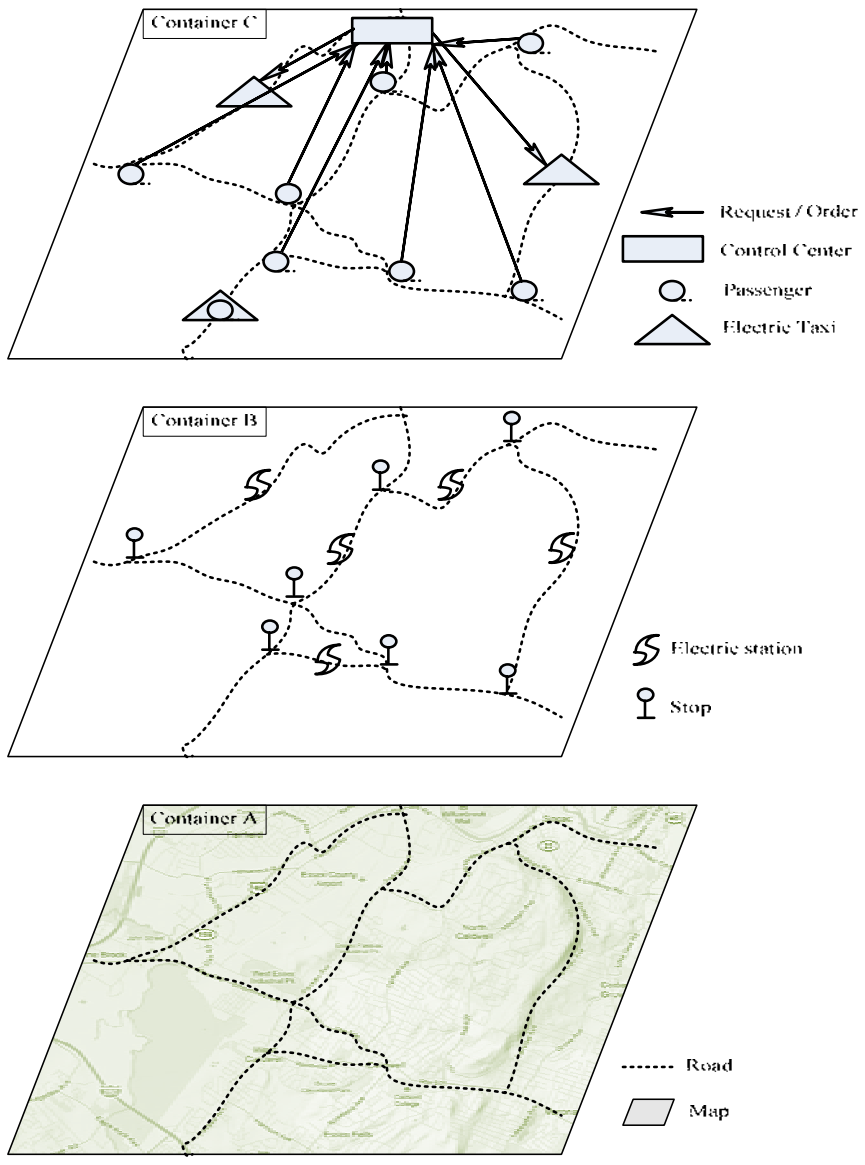

Fig. 2. The agent type of each container

The framework of the electric station agent is diagramed as figure 1. In the framework of electric station agent, there is one kind of information source: Electric-taxi agents. For electric station agent, it has two kinds of knowledge: The information about its' attributes and animation parameters. Combine the knowledge and information that perceived, and as the input data of decisions[4]. There is one decision rule: Update state and data. Through decision rules, taxi agent executes proper actions to one kind of agents: Electric-taxi agents. Finally, the spatial behavior of electric station agent will be displayed on the screen via animation parameters.

Next, we identify the interaction behaviors which in the interaction behaviors container of electric-taxi DAR operation system[5,6], and find out the basic interaction behaviors are consisted of three parts: First is traffic jam interaction behaviors, second is electricity replenishment interaction behaviors and third is DAR interaction behaviors. The structure of basic interaction behaviors is diagramed as figure 3.

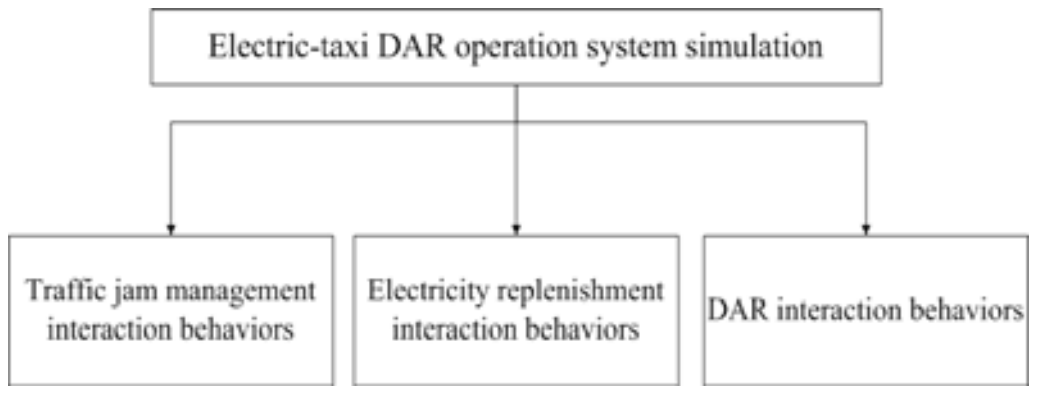

Fig. 3. The electric-taxi DAR operation system interaction behaviors

### 2.1 The Electric-taxi Agents' Interaction Behaviors

The three agents' interaction behaviors are: (1) Dial-a-ride and deliver interaction behaviors (2) traffic jam management interaction behaviors (3) electricity replenishing interaction behaviors.

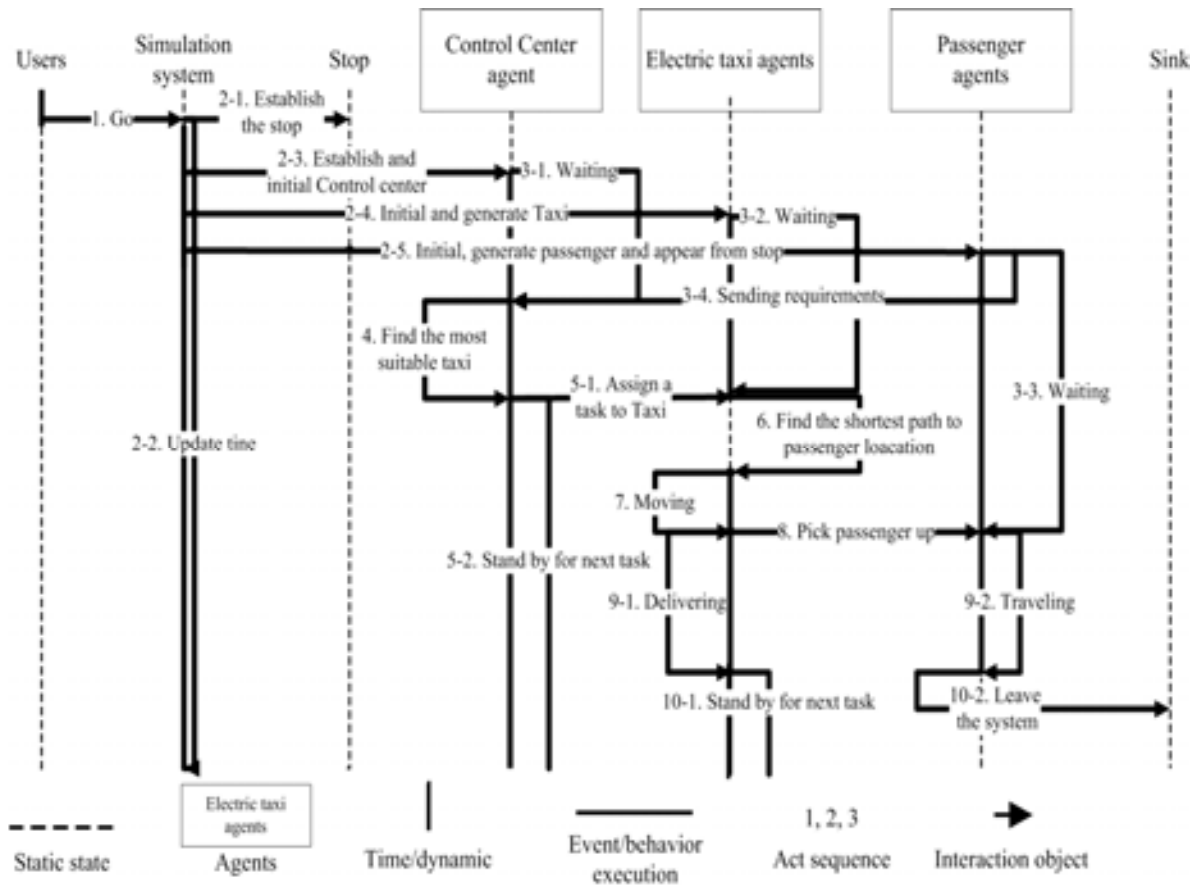

Fig. 4. Dial-a-ride interaction behaviors

Figure 4 is the dial-a-ride interaction behaviors diagram. In the electric-taxi agents' interaction behaviors diagram, a rectangle represents an actor that also means the executor of an event/behavior. The dotted line represents the state of actor is static, that means actor doesn't do any action. The straight thick line represents the state of actor is dynamic with time. The length of straight thick line represents how long the behaviors/event will last. The transverse thick line represents an event/behaviors is executed. The arrow of transverse thick line represents the interaction object of an event/behaviors execution. Final, the number which is in front of every event/behaviors represents the execution sequence.

Figure 5 is the illustration of electricity replenishment interactive spatial behaviors. In this electricity replenishment interaction behavior[7], there are three agents: Elec-tricity stations, control center and electric-taxis. The detail of behavior is described as below: In the beginning user execute the simulation (step1). Immediately, simulation initial and electricity stations, control center and electric-taxis (step2-2, 2-3 and 2-4), and simulation time is begin running and updating (step2-1). During the simulation period, electricity stations will update its' state (busy/idle) (step3). After generating and simulating for a while, electric-taxis can be divided into two kind of electric-taxis : Some electric-taxis are in replenishing and others are still in operation with electricity shortage (represented by the longest transverse thick-line). Electric-taxis those are in operation with electricity shortage ask electricity shortage information to control center and other electric-taxis those are in replenishing (step4-1 and 4-2). The electric-taxis that are in replenishing and control center to other electric-taxis those are in operation (step5-1 and 5-2). And then electric-taxis those are in lack of electricity select a most suitable electricity station according evaluation criteria (step6). After, electric-taxi moves to selected elected station (step7). When electric-taxis enter in or leave the selected electricity station, the electricity station will update its' state (busy/idle) (step8-1 and 10-1). When electric-taxi arrives at electricity station, driver has to waiting for replenishing until there are no cars in front of him (step8-2 and 9). Finally, after replenishing finished, electric-taxi will move to the closest stop and stand by for assignment (step10-2).

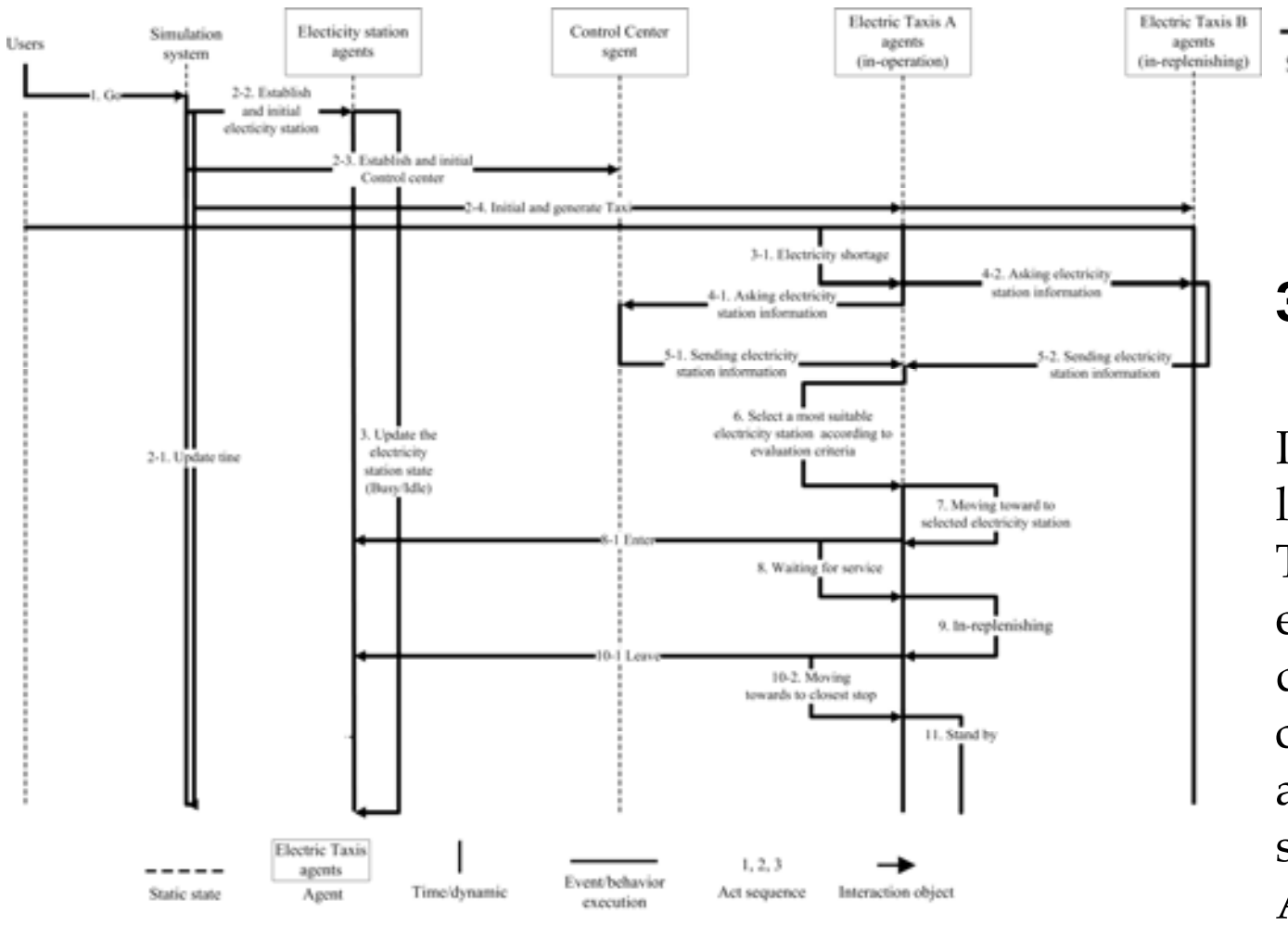

Fig. 5. Electricity replenishment interactivespatial behaviors

In this traffic jam management interactive spatial behaviors (Figure 6); there is one type of agents: Electric-taxis. The detail of behaviors is described as below: In the beginning user execute the simulation (step1). Immediately, simulation initial and generate two kind of actors (step2-2 and 2-3), and simulation time is begin running and updating (step2-1). During the simulation period, road will update its' state (traffic jam/ freely flowing) (step3). After generating and simulating for a while, electric-taxi can be divided into two kind of electric-taxis : Some electric-taxis are getting in traffic jam and others are moving freely (represented by the longest transverse thick-line). The electric-taxis those are getting in traffic jam send traffic jam information to each other and other electric-taxis those are not getting traffic jam (step4-2, 4-3 and 4-4). When electric-taxis enter in or leave road, the state of road will be changed into jam-packed or un-crowded (step4-1 and 6-1). When electric-taxi moves to a traffic jam road, driver will find another shortest path that excluded the traffic jam road (step5 and 6-2). And then electric-taxi will switch into another road and move on (step7). Finally, when electric-taxis that are leaving the traffic jam road, they will sending the traffic jam information again (step8).

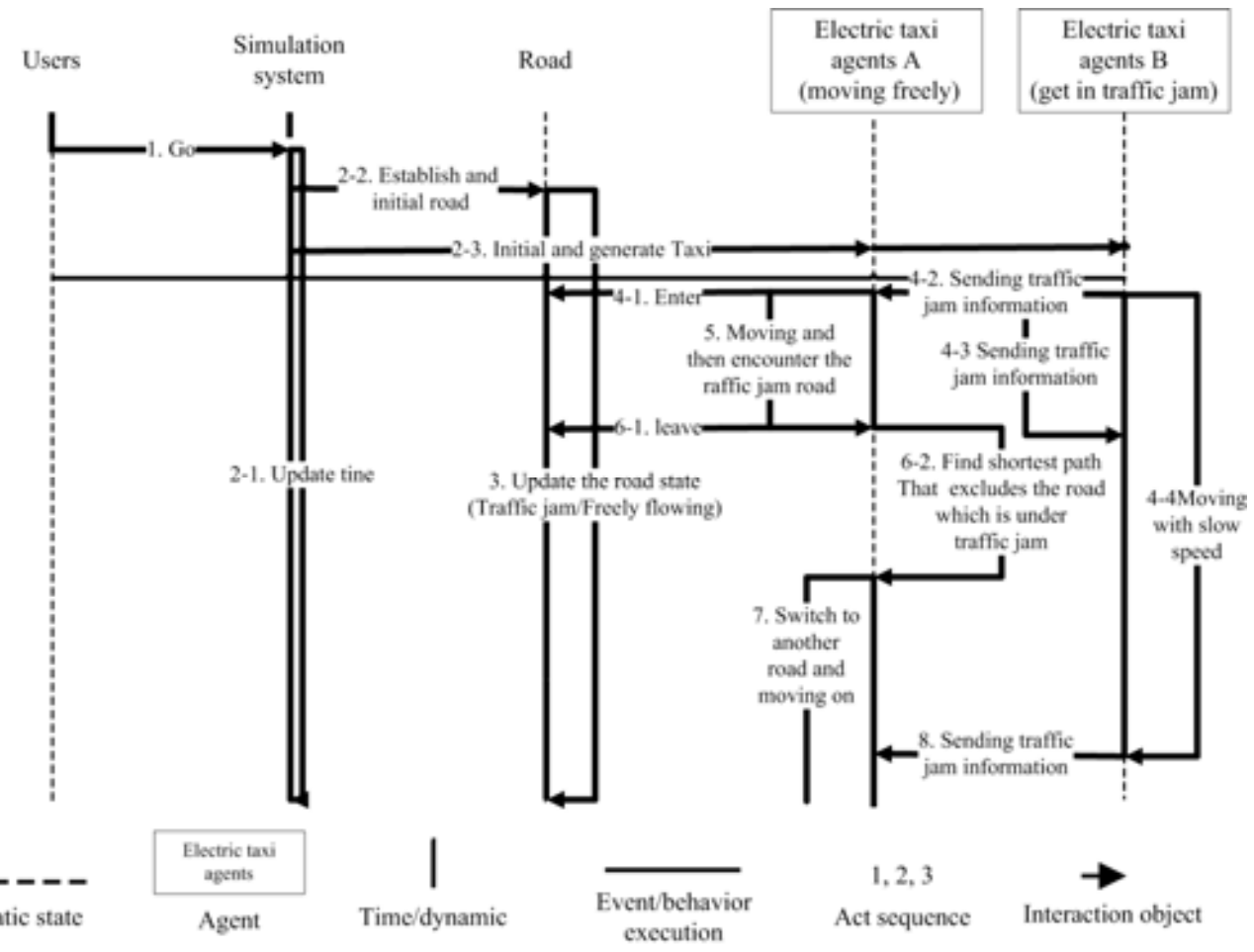

Fig. 6. Traffic jam interactive spatial behaviors

## 3 Simulation of Electric-Taix DAR Operation

In our simulation, we use java language as our program language. Java is an object-oriented program language[8]. That makes modeling of agent-based simulation become easier. And its' flexibility makes it can operate in many different platforms (for example: Linux, Unix, Mac, Microsoft Windows and so on). That's also its' the biggest advantage. On the other hand, AnyLogic is a java-based simulation platform, it just match our need. So we chose AnyLogic as our simulation platform.

According to the framework that proposed in the Section 2, we have constructed the multi-agent electric-taxi DAR operation system (figure 7).

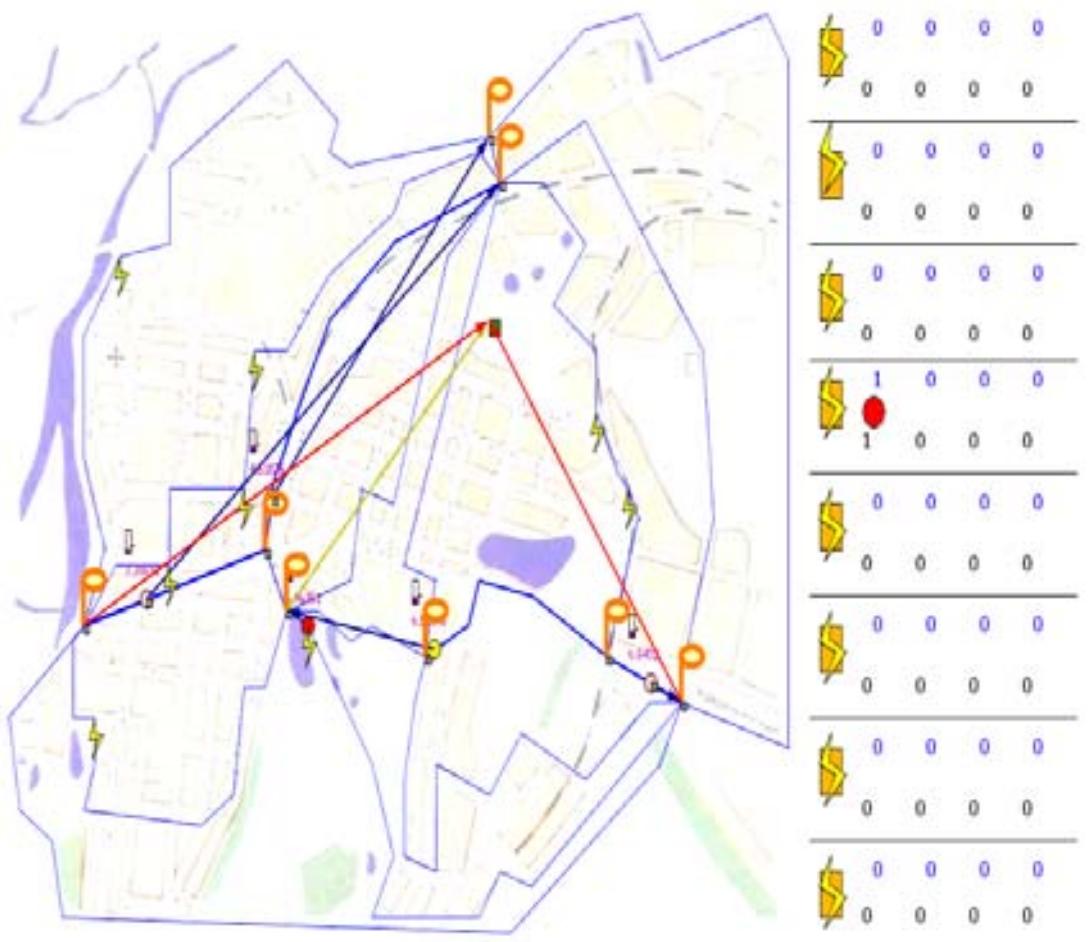

Fig. 7. The simulation of electric-taxi DAR operation system

Figure 8 illustrates the attributes of agent. The attributes of agent are extracted from the framework of agent.

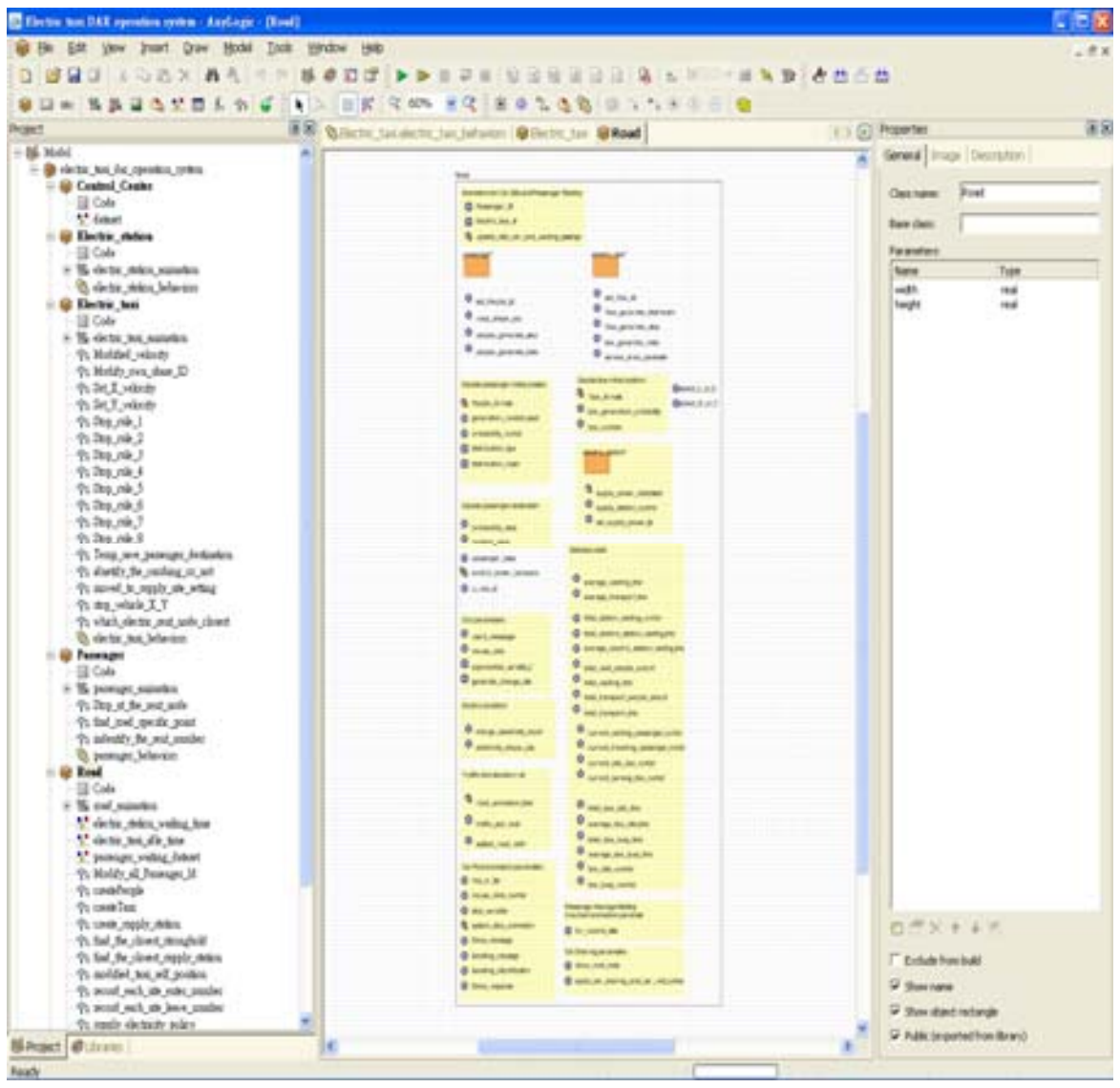

Fig. 8. The agent attributes of electric-taxi DAR operation system

## 4 Electric Taxi DAR Operation Case Study

In order to deal with all of issues mentioned above, we have to construct an electric taxi DAR operation system to simulate the operation of the electric taxi system in the city. And our performance measures are passenger average waiting time, electric taxi average idle time and electric taxi average waiting time in queue[9,10]. Then we will collect and analyze the data obtained from simulation. Finally, according to analysis results, we will make recommendations for the electric taxi company. The details of decision analysis for electric taxi DAR operation system simulation are described as follows.

### 4.1 Decision Analysis of Electric Taxi DAR Operation System

For an electric taxi company, it has its own electric taxi fleet. And we assume that it has 8 electric stations, no carpool and car-sharing management policies. Now it wants to know what the most feasible scale of electric taxi is. Because it thinks: "If the electric taxi scale is too large, the passenger waiting time will be lower but the electric taxi idle time will be higher during the rush hour. If the electric taxi scale is too small, the passenger waiting time will be higher but the electric taxi idle time will be lower during the rush hour."

In the downtown of city, the passenger density is higher than other places. And during the rush hour, if there are too many electric taxis on the same road, the traffic jam will happen. There are totally eight towns in the city.

First, we want to find out how large the scale of electric taxi fleet is feasible. So we fix three parameters: (1). Under rush hour condition. (2). Electric station number is 8. (3). Paths planning: In the beginning, we choose the shortest distance as our path planning[6]. In order to find out what's the feasible scale of electric taxi fleet, we vary the electric taxi number to observe the variances of passenger average waiting time and electric taxi average idle time.

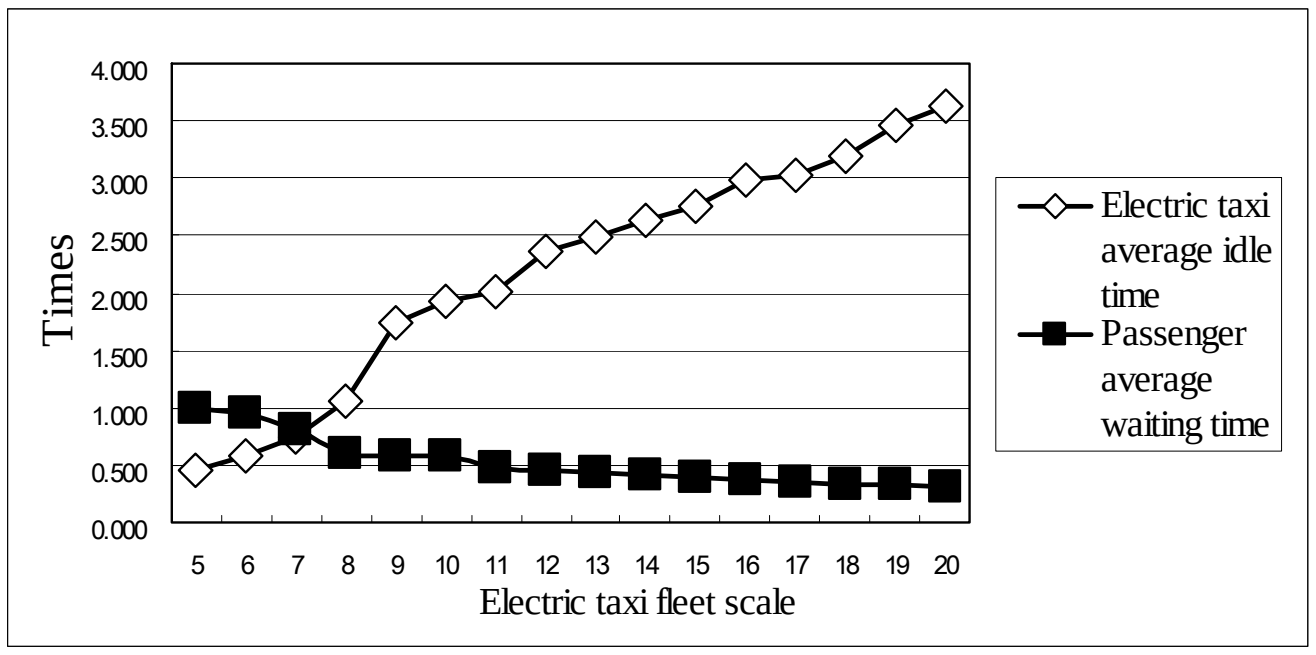

Fig. 9. Electric taxi fleet scale

From figure 9, we can obviously realize that if the scale of electric taxi fleet is less than 7, the passenger average waiting time is higher than others. Contrary, if the scale of electric taxi fleet is more than 7, the electric taxi average idle time is too high. The most important thing is what the most feasible scale of electric taxi fleet is. We decide the scale of electric taxi fleet is 11. The reason is that when electric taxi fleet scale is larger than or equals to 11, the passenger average waiting time begins converging.

Second, after deciding the scale of electric taxi fleet, the company wants to know how large the scale of electric station is feasible. If the scale of electric station is too small, the electric taxi waiting time will be too long. So we fix three parameters: (1). Under rush hour condition. (2). Electric taxi fleet number is 11. (3). Paths planning: we choose the shortest distance as our path planning. In order to find out what's the feasible scale of electric station, we vary the electric station scale from 1 to 8 to observe the variances of passenger average waiting time and electric taxi average waiting time in queue.

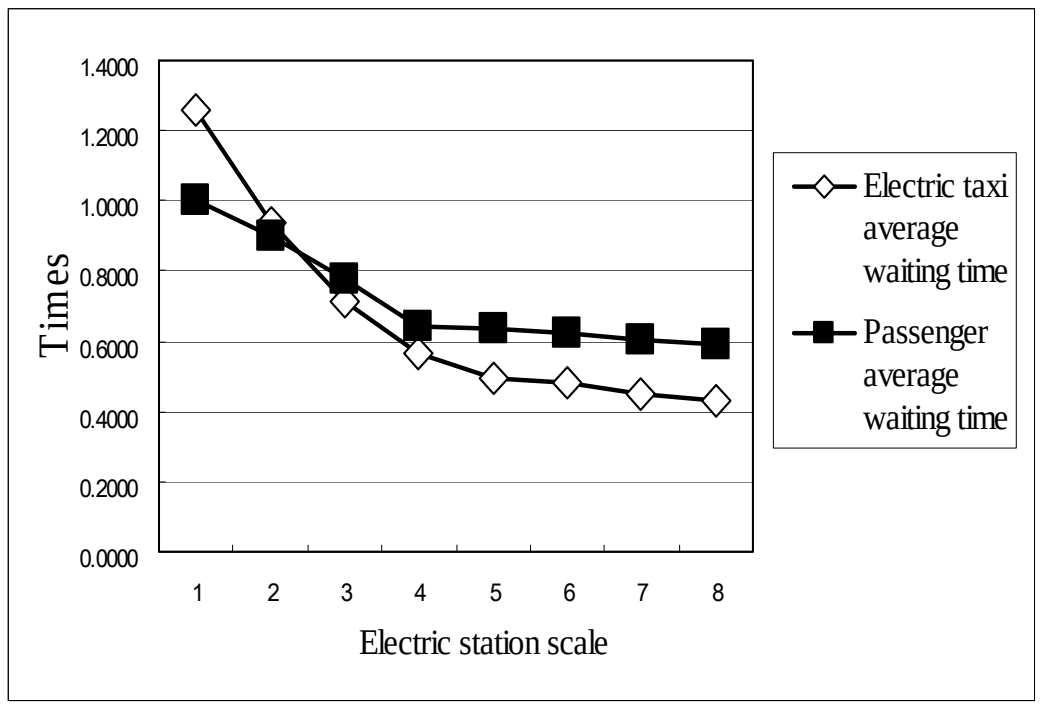

Fig. 10. Electric station scale

From figure 10, we can obviously realize that if the scale of electric station increases, both of the passenger average waiting time and electric taxi waiting time in queue decrease. Contrary, if the scale of electric station decreases, both of the passenger average waiting time and electric taxi waiting time increase. And we find out that if the waiting time of an electric taxi in queue is longer, the passenger has to wait for a long time. So the waiting time of an electric taxi is long, the passenger average waiting time is long. And the most important thing is what the most feasible scale of electric station is. We find out that if the scale of electric taxi fleet is equal to or large than 5, the average waiting time of an electric taxi and passenger average idle time begins converging. Hence, we decide the scale of electric station is 5.

Third, the company has decided the most feasible scale of electric taxi fleet and electric station, following the company wants to know which paths planning is better or more suitable for its' electric taxi fleet and electric taxi company operation. So we fix three parameters: (1). Under rush hour condition. (2). Paths planning: we choose the least time as our path planning. (3). Electric station number is 5. In order to find out which paths planning is better; we vary the electric taxi number from 5 to 20 to observe the variances of passenger average waiting time and electric taxi average idle time.

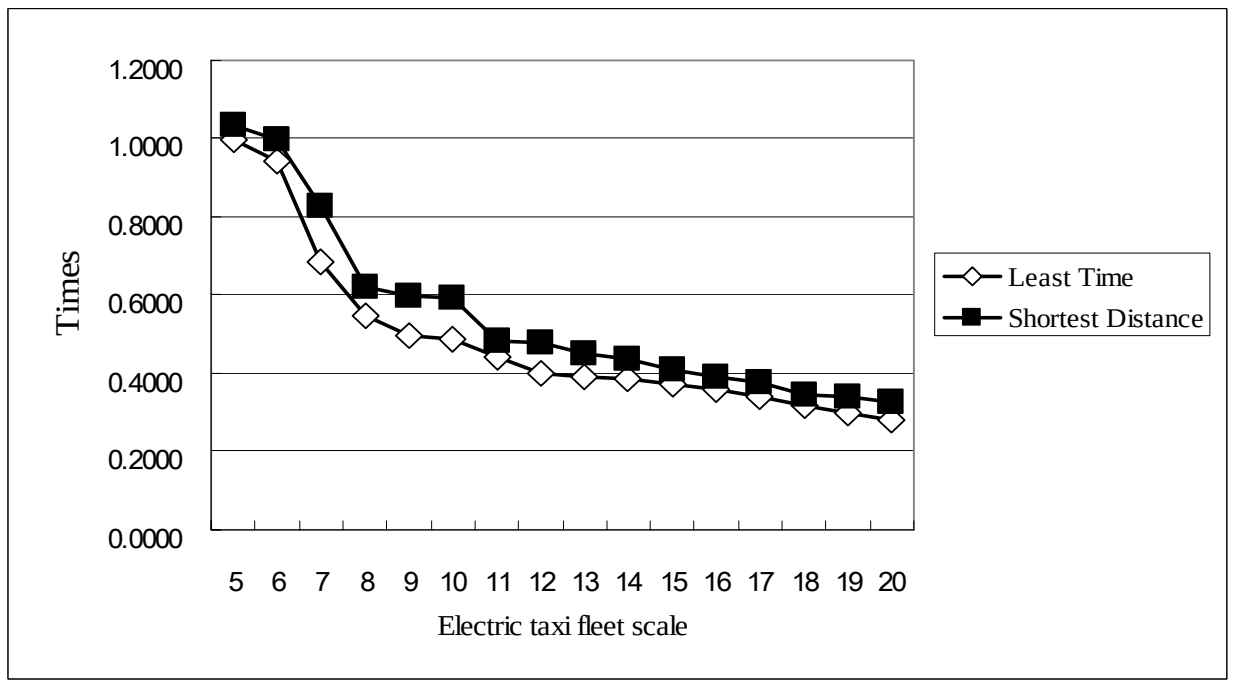

Fig. 11. Passenger average waiting time

From figure 11, we can clearly know all of the passenger average waiting time is better under least time path planning. And there is a trend in figure 11. First, if the scale of electric taxi fleet increases, the passenger average waiting time will decrease. Contrary, if the scale of electric taxi fleet decreases, the passenger average waiting time will increase. Second, we find out that if the scale of electric taxi fleet is equal to or large than 11, the passenger average waiting time begins converging. In brief, the company adds more electric taxi in the electric taxi fleet that only decreases small passenger average waiting time.

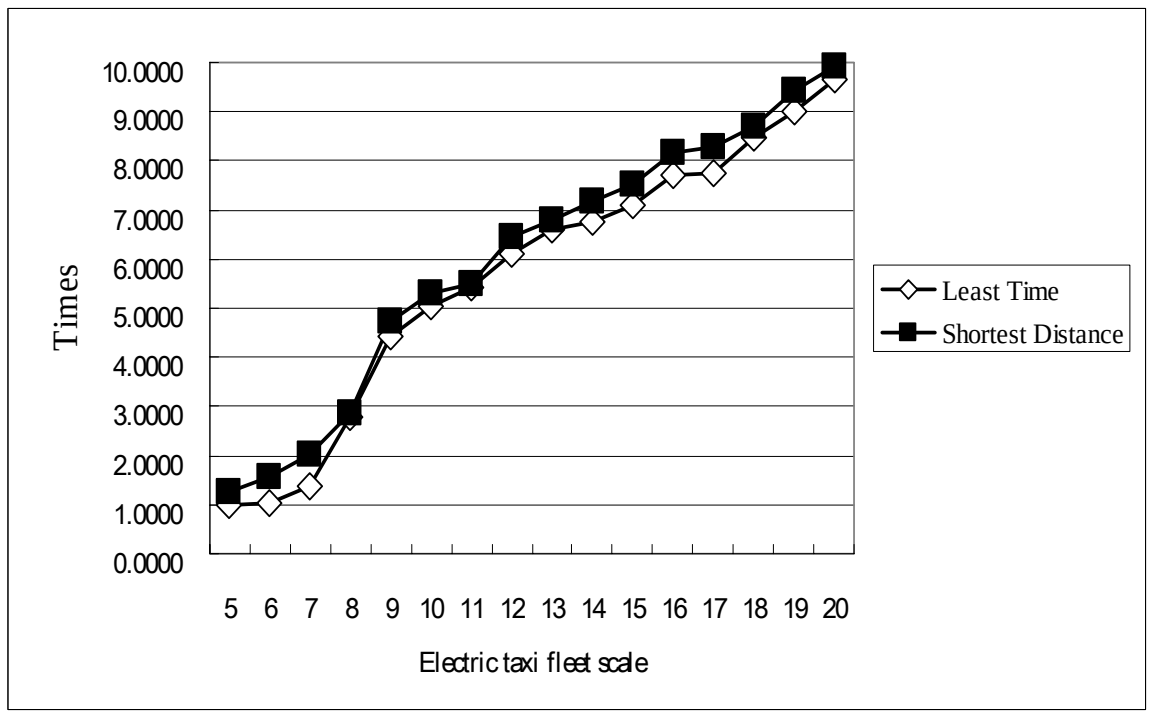

Fig. 12. Electric taxi average idle time

In figure 12, we can clearly know most of the electric taxi average idle time is worse under least time path planning. If the scale of electric taxi fleet increases, the electric taxi average idle time also increases. Contrary, if the scale of electric taxi fleet decreases, the electric taxi average idle time will decrease. So we can make a brief summary: The electric taxi company applies least time as path planning is more suitable for electric taxi fleet and passengers. Even though the electric taxi fleet scale increases causes the electric taxi average idle time significantly increases, it also represents that the small scale of electric taxi fleet can achieve the lower passenger average waiting time.

### 4.2 Decision Analysis of Car-Pool and Car-Sharing Management Policies

After electric taxi company has determined the electric taxi fleet scale, electric station scale and path planning. The electric taxi company still want to add car-pool management policy into the electric taxi operation[11]. Nevertheless, they have to know the performance of car-pool to measure the feasibility.

So we fix four parameters: (1). Under rush hour condition. (2). Paths planning: we choose the least time as our path planning. (3). Electric station number is 5. (4). Car-pool. In order to find out whether apply car-pool is better or not; we vary the electric taxi fleet scale from 5 to 20 to observe the variances of passenger average waiting time and electric taxi average idle time.

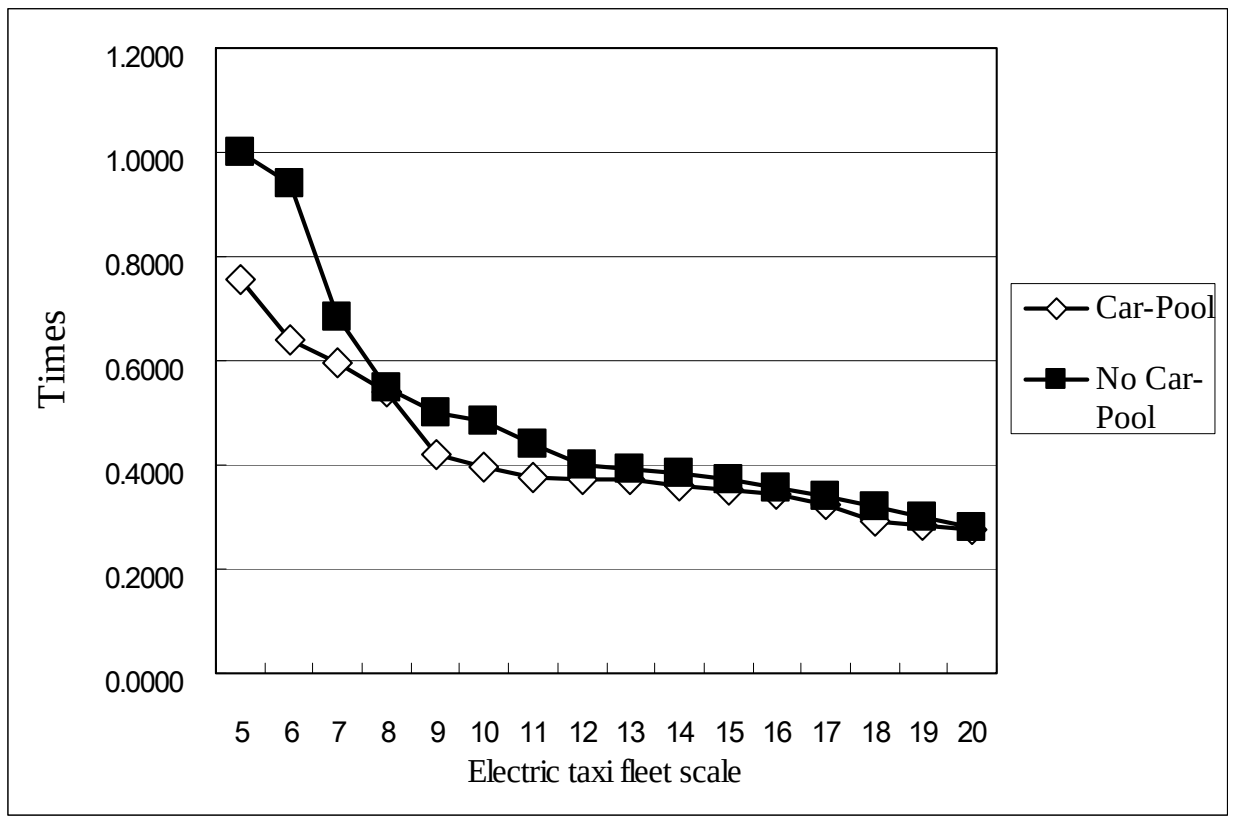

Fig. 13. Passenger average waiting time

From figure 13, we can clearly know all of the passenger average waiting time is lower under Car-Pool. First, if the scale of electric taxi fleet increases, the passenger average waiting time will decrease. Contrary, if the scale of electric taxi fleet decreases, the passenger average waiting time will increase. Second, we find out that if the scale of electric taxi fleet is equal to or large than 11, the passenger average waiting time begins converging. In brief, the company adds more electric taxi in the electric taxi fleet that only decreases small passenger average waiting time.

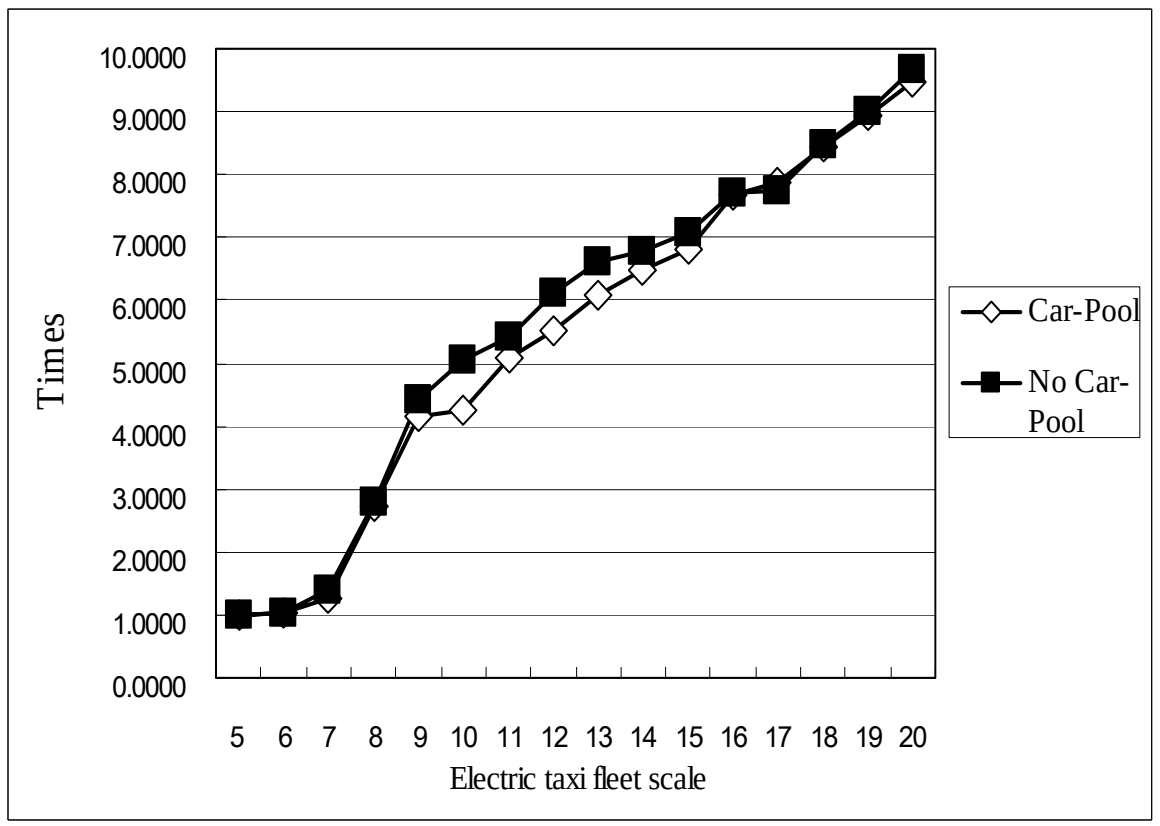

Fig. 14. Electric taxi average idle time

In figure 14, we can clearly know all of the electric taxi average idle time is better under Car-Pool. If the scale of electric taxi fleet increases, the electric taxi average idle time will increase. Contrary, if the scale of electric taxi fleet decreases, the electric taxi average idle time will decrease. So we can make a brief summary: The electric taxi company applies car-pool is more suitable for electric taxi fleet and passengers. So we can make a conclusion for this discovery: The electric taxi company just need own small electric taxis fleet, they can obtain the same or better the result of performance measure as electric taxi company owns large electric taxi fleet scale. After understanding the data analysis, the electric taxi company is willing to practice the Car-Pool management policy in the electric taxi fleet operation.

So far, the electric taxi company satisfies its operation performance. In addition to car-pool, they also want to evaluate the effect of car-sharing[12]. This management policy means the electric taxi company has to divide their electric taxi fleet into two types: (1). Normal electric taxi fleet: an electric taxi which does anything we mentioned before. (2). Rental car: passengers only pay little fees to electric company, they can drive a car to their destinations where has the site of getting and return. Before applying car-sharing management policy, the electric taxi company needs to consider two important things: (1). the rental car comes from the original electric taxi fleet. So the scale of electric taxi fleet becomes smaller than before, the service level and efficiency may be debased. (2). Apply the car-sharing will get a better performance?

Hence we fix five parameters to do the simulation: (1). Under rush hour condition. (2). Paths planning: we choose the least time as our path planning. (3). Electric station number is 5. (4). Car-pool. (5). Car-sharing: we do apply car-sharing. In order to find out whether apply car-sharing is better or not; we vary the electric taxi number from 5 to 20 to observe the variances of passenger average waiting time and electric taxi average idle time.

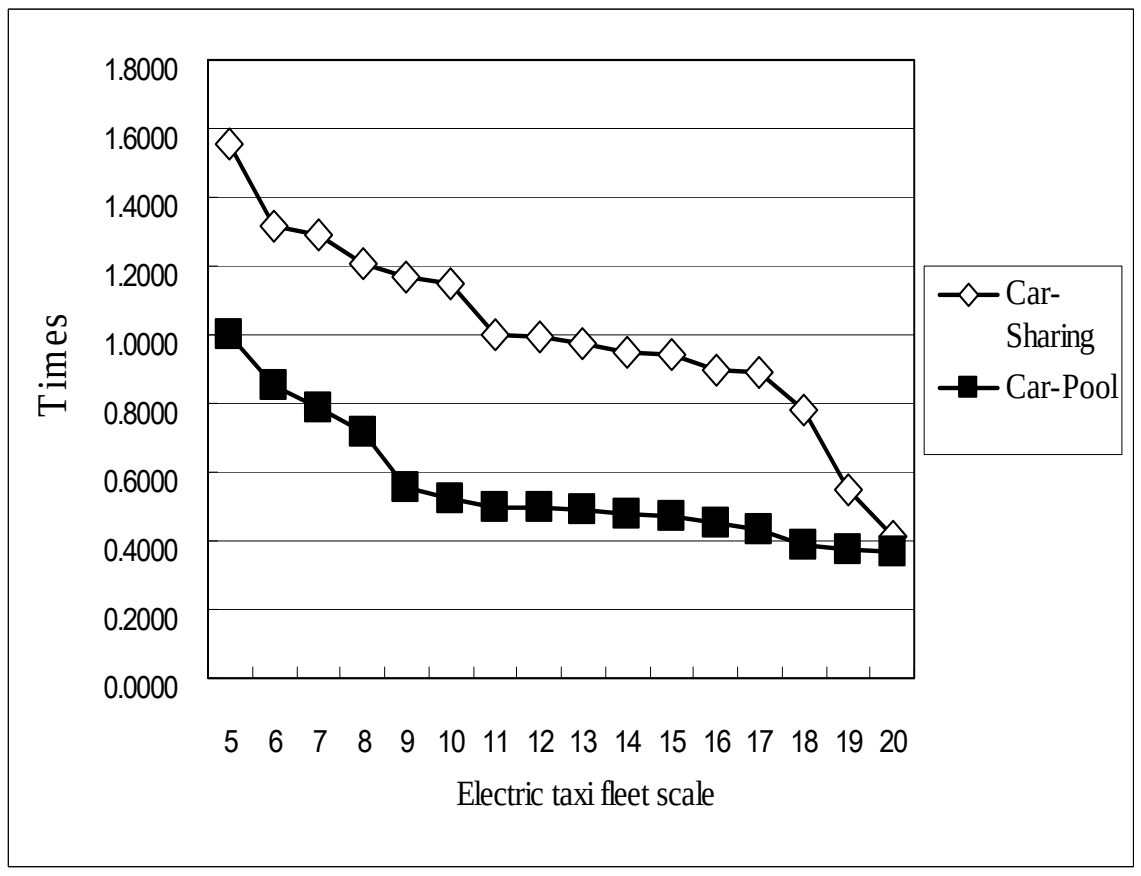

Fig. 15. Passenger average waiting time

From figure 15, we can clearly know all of the passenger average waiting time is higher under Car-Sharing. First, if the scale of electric taxi fleet increases, the passenger average waiting time will decrease. Contrary, if the scale of electric taxi fleet decreases, the passenger average waiting time will increase. Second, under car-pool, we find out that if the scale of electric taxi fleet is equal to or large than 11, the passenger average waiting time begins converging. In brief, the company adds more electric taxi in the electric taxi fleet that only decreases small passenger average waiting time but increase more electric taxi idle time. On the other hand, from figure 15, we find out that divides the original electric taxi fleet into two types can decrease the service level of electric taxi fleet. In other words, due to the reduction of scale for electric taxi fleet, the electric taxi number of satisfying passenger demands also becomes less.

From figure 15, we can find out that electric taxi average idle time under Car-Sharing is also worse than Car-Pool. And we realize that if we apply car-sharing in our electric taxi operation, we will get a worse electric taxi average idle time. That's an interesting phenomenon. We remain the same scale of total service car; we just divide the original electric taxi fleet into two types. But rental cars only serve two places. Other electric taxis serve the whole city. If one specific place doesn't appear any passenger for a long time (because the passenger generation belongs to normal distribution), the rental cars will still wait until a passenger appears. We can image that there are many rental cars in the same place, but there is only one passenger shows up. Under this condition, only one car will be used and other cars still wait for next passenger. Therefore, we can conjecture that the electric taxi average idle time will increase.

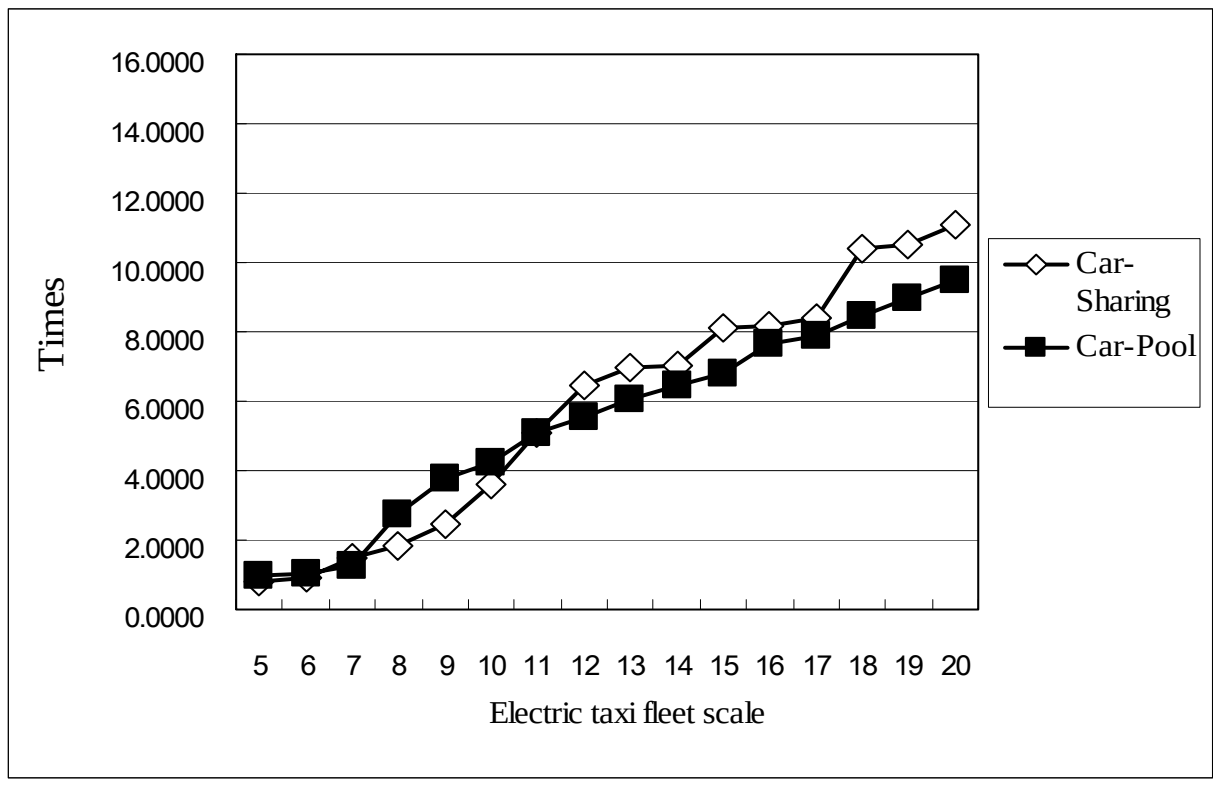

Fig. 16. Electric taxi average idle time

From 16, we can clearly know all of the electric taxi average idle time is worse under Car-Sharing. If the scale of electric taxi fleet increases, the electric taxi average idle time will increase. Contrary, if the scale of electric taxi fleet decreases, the electric taxi average idle time will decrease. From figure 16 and figure 17, we can make a brief summary: When the electric taxi number is from 5 to 9, the passenger demands is still much more than electric taxi fleet scale and rental cars. Plus rental cars are divided from the original electric taxi fleet that causes reduction of the electric taxi fleet scale. Even the electric taxi average idle time is still increasing under car-sharing; it is smaller than the value of Car-Pool.

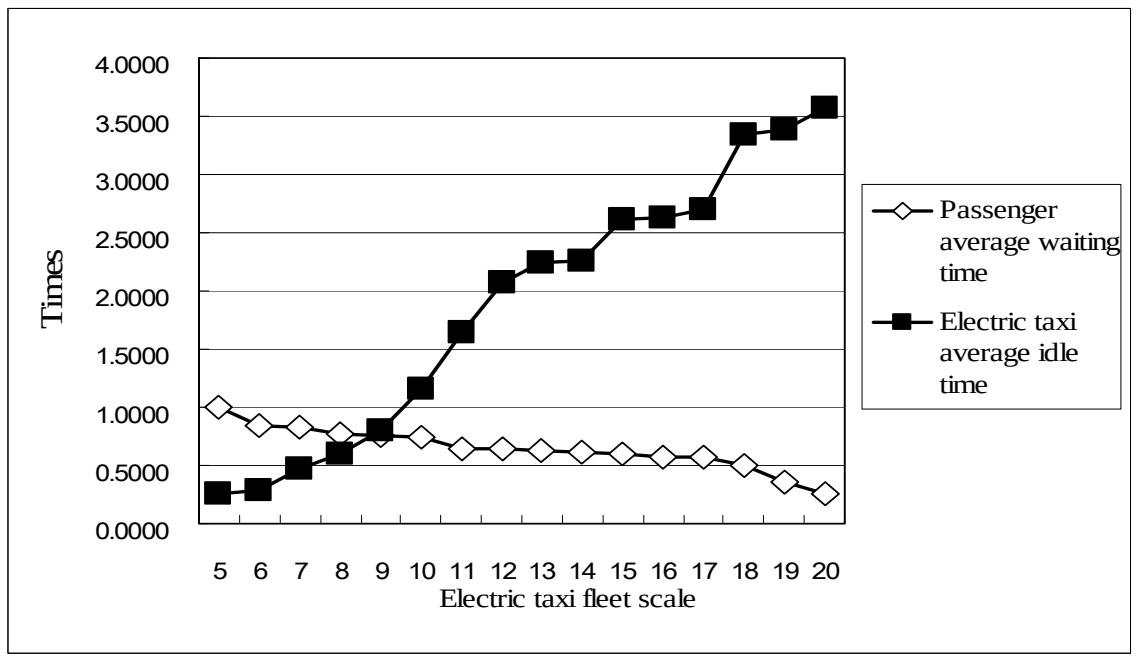

Fig. 17. Car-sharing situation

Finally, we make a summary. Because the original electric taxi fleet is divided into two types that cause the scale of the electric taxi fleet becomes smaller than before. Even the rental car is more convenient and faster for specific passenger type, for the electric company, they will not have enough electric taxi number to satisfy other numerous passengers. Consequently, the performance measure won't correspond with our mission. Obviously, introducing car-sharing management policy will negatively affect the electric taxi operation and performance. After data analysis, the electric company will stop introduce car-sharing management policy temporarily and do the further assessment.

## 5 Conclusions

In our research, we not only construct a multi-agent simulation framework. In order to approach reality, we consider the change of electric taxi speed, electric taxi power, traffic jam simulation management and so on to make our multi-agent simulation be more credible and reliable. In addition, we provide three kinds of management policies (paths planning, car-pool and car-sharing) to observe the feasibility and efficiency to provide supervisor of taxi company as their decision-making suggestions. Following, we list the contribution that we has achieved:

First, we apply three management policies (paths planning, car-pool and car-sharing) into the electric taxi DAR operation system to observe the phenomenon and effects. We find out that apply least time as our path planning and car-pool to manage the electric taxi DAR operation system, the performance of electric taxi fleet (passenger average waiting time and electric taxi idle time) can achieve our standard. But add car-sharing into the electric taxi DAR operation system will cause the reduction of electric taxi fleet performance. So car-sharing management policy affects, even reduce the efficiency and utility of car-pool, if they are implemented in the electric taxi DAR operation system in turn.

Second, we have constructed a graphic user interface (GUI) for user. User can control the passenger generation rate, electric taxi fleet scale and electric station number to fit into the operation of real taxi company via our GUI. User also can make up any combination of management policies (Paths Planning, Car-Pool and Car-Sharing) to observe the result of simulation. According to the simulation result, the taxi company can modify their management policy to be more competitive and quicker response with the real condition. Our GUI also can display the electric taxi idle and passenger waiting message to let taxi company supervisor control the condition real time. We also offer a communication window between agents and user, to support the user intervene the simulation and force the agents need to interact with user to negotiate with passenger.

## Acknowledgment

This research work was sponsored by the National Science Council, R.O.C., under project number NSC98-2221-E-155-023.

**Yee Ming Chen** is a professor in the Department of Industrial Engineering and Management at Yuan Ze University, where he carries out basic and applied research in agent-based computing. His current research interests include soft computing, supply chain management, and pattern recognition.

**Bo-Yuan Wang** is a graduated student in the Department of Industrial Engineering and Management at Yuan Ze University,, where he is studying basic and applied research in agent-based programming. His current research interests include mathematical programming computation and system analysis/synthesis.

**Hung-Ming Shiu** is a graduated student in the Department of Industrial Engineering and Management at Yuan Ze University,, where he is studying basic and applied research in multi-agent simulation. His current research interests include web-based programming and software system analysis.